\magnification=1200
\medskip
\centerline {\bf CHERN-SIMONS SOLITONS IN QUANTUM POTENTIAL}
\medskip
\bigskip\bigskip
\centerline{O. K. PASHAEV{$^{1,2}$}\footnote\dag{e-mail:
pashaev@math.sinica.edu.tw; pashaev@vxjinr.jinr.ru} and Jyh-Hao LEE{$^{2}$}\footnote{$^\star$}{e-mail: leejh@ccvax.sinica.edu.tw;
Fax: 886-2-27827432}}
\bigskip
\centerline{$^{1}$Joint Institute for Nuclear Research, 
Dubna (Moscow), 141980, Russia}
\bigskip
\par
\centerline{$^{2}$Institute of Mathematics, Academia Sinica, 
Nankang, Taipei 11529, Taiwan}
\bigskip
\medskip
\bigskip\par\noindent
{\bf Abstract}---{\sl The self-dual Chern-Simons solitons under
the influence of the quantum potential are considered.
The single-valuedness condition for an arbitrary integer number 
$N \ge 0$ of solitons leads to quantization
of Chern-Simons coupling constant $\kappa = m {e^{2} \over g}$, 
and the integer strength of quantum potential $s = 1 - m^{2}$. 
As we show, the Jackiw-Pi model corresponds to the first member 
(m = 1) of our hierarchy of the Chern-Simons
gauged nonlinear Schr\"odinger models, admitting self-dual 
solitons. New type of exponentially 
localized Chern-Simons solitons for the Bloch electrons near 
the hyperbolic energy band boundary are found.}
\bigskip\bigskip
\centerline{1. INTRODUCTION}
\bigskip\bigskip\par\noindent
 The Chern-Simons theory attracted much attention recently  
as a gauge theory in 2+1 dimensions [1], describing non-local
interaction between matter particles, affecting the phase 
of the wave function, and  leading  to the  fractional 
statistics phenomena [2]. In this connection the Chern-Simons 
coupling
constant plays the role of a statistical parameter and 
relates to the spin of particles.
At the classical non-relativistic level, with the Chern-Simons 
coupling strength  determined by overall strength of a quartic 
scalar potential,
it leads to the existence of the self-dual Chern-Simons 
solitons [3].  They are
static solutions with finite charge and flux
of exactly solvable self-dual equations [4]. Unfortunately, 
the full dynamics
of Chern-Simons solitons according to the nonlinear 
Schr\"odinger equation is still inflexible due to the 
lack of exact integrability, even without external forces,
although it may be realized in the framework of the 
Davey-Stewartson-II equation [5]. 
\par In the present paper we show that the "external" force
produced by the so-called "quantum potential"  
$U(x) = (- \hbar^{2}/2m)\Delta |\psi|/|\psi|$, leads to 
an additional
statistical transmutation of Chern-Simons solitons. 
The quantum
potential, introduced by L. de Broglie [6] and explored 
by D. Bohm [7]
does not depend on the strength of the wave but only on 
its form,
and therefore its effect could be large even at long distances.
Then, it satisfies the homogeneity property [8], this is the 
reason why it appears in attempts
of a nonlinear extension of the quantum mechanics [9,11-14].
An approach has been
developed to the stochastic formulation of quantum mechanics, 
where
the quantum fluctuations are represented by superimposing
a classical trajectory and an additional random motion 
generated by
quantum potential [10]. The nonlinear extension of the 
Schr\"odinger equation with the quantum potential non-linearity 
has been considered in connection with
several problems [11]: a) in allowing formally the 
diffusion coefficient of the stochastic process in a 
stochastic quantization  to differ $\hbar /2m$, related 
to the difference in the Plank constant [12] or the 
inertial mass
[13], b) in corrections from quantum gravity [14]. 
As was shown by Sabatier,
depending from intensity of the quantum potential it can be 
linearized in the form of the Schr\"odinger equation with 
rescaled potential or as the pair of time reversed diffusion 
equations [15], and in both cases does not admit 
soliton solutions.
Recently we find that for self-consistent potential 
$U = g|\psi|^{2}$
in 1+1 dimensions the theory has soliton solutions 
with rich resonance dynamics [16]. In the present paper 
we find exactly soluble case for
2+1 dimensional nonlinear Schr\"odinger (NLS) model 
interacting with Chern-Simons
field under the influence of the quantum potential.
\par
In the above mentioned interpretation of quantum mechanics 
the quantum particle moves
under the action of a force which along with classical potential
includes a contribution from the quantum potential.
If instead of classical particle we consider
Chern-Simons solitons, then subject
to the influence of intensity $s$ quantum potential, 
it could represent
the stochastically quantized anyons.  In this case 
we can expect
quantization condition on the Chern-Simons coupling 
constant,
an affect on the anyonic parameter and the appearance of 
the zero point fluctuation for the statistical flux.
\bigskip\bigskip
\centerline{2. CHERN-SIMONS SOLITONS HIERARCHY} 
\bigskip\bigskip\par
We consider the Chern-Simons gauged Nonlinear 
Schr\"odinger model
(the Jackiw-Pi model) with nonlinear quantum potential term
of strength $s$:
$$
L = {\kappa \over 2}\epsilon^{\mu\nu\lambda}A_{\mu}
\partial_{\nu}A_{\lambda}
 + {i \over 2}(\bar\psi D_{0}\psi - \psi \bar D_{0}\bar \psi)
- \bar{\bf D}\bar\psi {\bf D}\psi + 
s\nabla |\psi| \nabla |\psi|
+ g |\psi|^{4}, \eqno(1) $$
 where $D_{\mu} = \partial_{\mu} + ieA_{\mu}$, 
$(\mu = 0,1,2)$. Classical equations
of motion are
$$iD_{0}\psi + {\bf D}^{2}\psi + 2g |\psi|^{2}\psi = 
s {\Delta |\psi|
\over |\psi|}\psi, \eqno(2a)$$
$$\partial_{1}A_{2} - \partial_{2}A_{1} = 
{e \over \kappa}\bar\psi\psi,
\eqno(2b)$$
$$\partial_{0}A_{j} - \partial_{j}A_{0} = 
-{e \over \kappa}i\epsilon_{jk}
(\bar\psi D_{k}\psi - \psi \bar D_{k}\bar\psi),\,\, 
(j,k = 1,2).
\eqno(2c)$$
As we mentioned in Introduction the extension of the 
linear Schr\"odinger equation  with $s \ne 0$ term 
has been considered before [11-14], and was linearized
for {\it any} $s$ [12]. But, when one adds the 
nonlinear self-interaction and the gauge field,
integrability of the model is allowed only under 
some restrictions on
the coupling parameters.
For $s < 1$, and this is only the case where we 
find the self-duality
condition, we decompose $\psi = e^{R - iS}$ and 
introduce new rescaled variables
$$t = (1 - s)^{1/2} \tilde t,\,\,\,\,S = 
(1 - s)^{1/2}\tilde S, $$
$$A_{0} = (1 - s)\tilde A_{0}, \,\,\, {\bf A} =
(1 - s)^{1/2}\tilde {\bf A}.\eqno(3)$$
Then for new function $\tilde \psi = 
\exp (R - i\tilde S)$,
we get the gauged NLS system (the Jackiw-Pi model)
$$i\tilde D_{\tilde 0}\tilde\psi + 
\tilde{\bf D}^{2}\tilde\psi + 
{2g \over 1 - s}|\tilde\psi|^{2}\tilde\psi = 0, \eqno(4a)$$
$$\partial_{1}\tilde A_{2} - 
\partial_{2}\tilde A_{1} = {e \over \kappa (1 - s)^{1/2}}\bar{\tilde\psi}\tilde\psi, \eqno(4b)$$
$$\partial_{\tilde 0}\tilde A_{j} - 
\partial_{j}\tilde A_{0} = 
-{e \over \kappa (1 - s)^{1/2}}i\epsilon_{jk}
(\bar{\tilde\psi} \tilde D_{k}\tilde\psi - 
\tilde\psi \bar {\tilde D_{k}}\bar{\tilde\psi}),
\,\, (j,k = 1,2),
\eqno(4c)$$
but with new coupling constants $\tilde \kappa = 
\kappa (1 - s)^{1/2}$,
$\tilde g = g /(1 - s)$.
Since invariance of the density $\rho = \bar\psi \psi = \tilde{\bar\psi}\tilde\psi$, the Gauss law (4b) 
modifies the value
of magnetic flux, which depends now of $s$.
Only when the coupling constants are restricted 
by the condition
$$ {g\kappa \over e^{2}} = (1 - s)^{1 \over 2},\eqno (5)$$
static solutions of the system (4)
with the gauge potential
$$\tilde A_{0} = {e \over \kappa (1 - s)^{1/2}} 
\bar\psi\psi,$$
require the self-dual Chern-Simons equations,
$$\tilde D_{-}\tilde\psi = 0, \eqno(6a)$$
$$\partial_{1}\tilde A_{2} - \partial_{2} \tilde A_{1} =
{e \over \kappa (1 - s)^{1/2}}\tilde{\bar\psi}\tilde \psi. 
\eqno(6b)$$
Condition (5) extends for $s \ne 0$, the well-known  
Jackiw-Pi constraint. Namely, self-duality equations 
in the presence of
quantum potential survive  only if its strength is 
restricted by
 $s = 1 - g^{2}\kappa^{2}/e^{4}$.
\par To solve the system (6) we insert ${\bf A}$ 
from the first
equation to the second one and will get for the density
$\rho = \bar\psi \psi $ the Liouville equation
$$\Delta \ln \rho = - 2{e^{2} \over \kappa (1 - s)^{1/2}} \rho. 
\eqno(7)$$
The radially symmetric solutions, which have been constructed
by G. W. Walker [17], and discussed in [3],  
$$ \rho (r) = 4 {\kappa (1 - s)^{1/2} N^{2} \over e^{2} r^{2}}
[({r \over r_{0}})^{N} + ({r_{0} \over r})^{N}]^{-2}, \eqno(8)$$
would be regular if $N \ge 1$. Then,
like in [3], from regularity
of the gauge potential ${\bf A}$ we can
fix the phase of $\tilde \psi = 
\exp (R - i\tilde S)$
as $\tilde S = (N - 1)\theta$, $\theta = 
\arctan {x_{2}/x_{1}}$, and
restrict $N$ to be an integer for single-valued 
$\tilde \psi$.
However, the auxiliary function $\tilde \psi$ 
is not the physical one, 
this is why fixing an integer $N$ is not 
sufficient for single
valuedness of the original function 
$\psi = \exp (R - i(1 - s)^{1/2}
(1 - N)\theta)$. It would be now that the 
phase is restricted as
$0 < S \le 2\pi (1 - s)^{1/2}$, and in general 
we have the problem with
angular defect in the plane $2\pi (1 - (1 - s)^{1/2})$, 
describing a cone	. 
It is easy to see to avoid this complication
an integer valued must be the product
$$(1 - s)^{1/2}(N - 1) = n, \eqno(9)$$
valuedness of which for any integer $N$, 
requires an integer valuedness of
$$(1 - s)^{1/2} = m, \eqno(10)$$
and as the consequence of (5), we obtain the 
quantization condition
$${g\kappa^{2} \over e^{2}} = m,\,\,\,\,\, (m = 1,2,3...).
\eqno(11)$$
The last one means that in the presence of quartic 
self-interaction
the Chern-Simons coupling constant and the quantum 
potential strength
must be quantized
$$ \kappa = m {e^{2} \over g},\,\,\,\,\,\, s = 1 - m^{2}. 
\eqno(12)$$
When $m = 1$, the quantum potential vanishes, $s = 0$, 
while
the first constraint in (12) reduces to the Jackiw-Pi 
self-dual condition.
It is worth to note that in our relation (12) $g$ 
and $\kappa$ coupling constants
play the role similar to ones in 
the electro-magnetic duality condition derived
by Dirac. 
\par
The self-dual system (6) for original $(\psi,{\bf A})$
has the form, depending of the quantum 
potential with a coefficient proportional to the angular defect,
$$D_{-}\psi + ((1 - s)^{1/2} - 1){\partial_{-}|\psi| \over
|\psi|}\psi = 0, \eqno(13a)$$
$$\partial_{1} A_{2} - \partial_{2} A_{1} =
{e \over \kappa}\bar\psi \psi, \eqno(13b)$$
and turn the energy
$$H = \int d^{2}{\bf r}(\bar {\bf D}\bar\psi {\bf D}\psi -
s \nabla |\psi| \nabla |\psi| - g |\psi|^{4}), \eqno(14)$$
to vanish $H = 0$. It shows that under 
the influence of the quantum potential 
Chern-Simons solitons continue to be 
a zero energy configuration.
Due to (9), and (10) the flux for solution (8) is
quantized
$$\Phi = \int d^{2}{\bf r} B = 
{2\pi \over e} 2N (1 - s)^{1/2} =
 {2\pi \over e} (2m) N,\eqno(15)$$  
where $N = 1, 2, 3, ...$ and value of $m$ is fixed. 
Thus, the magnetic flux of our vortex/soliton 
is an even multiple $m$ of the elementary flux quantum, 
which generalizes the Jackiw-Pi result 
related to the particular value $m = 1$. 
Here it is worth to note that the single-valuednes condition 
(9) may be satisfied also for particular 
real values for $s$ and $N$. 
Thus, if $(1 - s)^{1/2} = p/q$ is the rational number, 
then single-valuedness of $\psi$ is allowed only 
for the specific sequence of integers 
$N = 1, 1 + q, 1 + 2q, ...$
or $ N = 1 + q l$, where $l = 0,1,2...$. 
In this case the flux is quantized
as 
$$ \Phi =  {2\pi \over e} (2p)(l + {1 \over q}),\eqno(16)$$
with $1/q$ playing the role of the  zero-point flux. 
For an irrational value $(1 - s)^{1/2} = \alpha$ 
the flux is quantized 
$$ \Phi =  {2\pi \over e} 2(n + \alpha),\,\,\, 
(n = 0,1,2,...),\eqno(17)$$
only for the irrational sequence $N = 1 + n/\alpha$.
\par
The decomposition $\psi = \exp (R - iS)$ is known in 
quantum mechanics as the Madelung fluid representation 
and has been explored for description 
of superconductivity [18]. In the problem (2) 
the velocity of associated Madelung fluid 
is definite as ${\bf v} = -2(\nabla S - e{\bf A})$ 
and satisfies the conservation law 
$$\partial_{t}\rho + \nabla \rho {\bf v} = 0.\eqno(18)$$
For the self-dual flows, from Eq.(6a) 
we find the canonical Hamiltonian equations 
$$ \dot x \equiv v_{1} = 
{\partial \chi \over \partial y},\,\,\,
   \dot y \equiv v_{2} = -{\partial \chi \over \partial x},
\eqno(19) $$ 
where the stream function $\chi$, 
playing the role of a Hamiltonian, 
has the form
$$\chi = (1 - s)^{1/2} \ln \rho.\eqno(20)$$
For the soliton solution (8), function $\rho$ 
is regular everywhere and
has $2(N - 1)$ -th order zero at 
the beginning of coordinates. This zero
determines singularity of $\chi$ and ${\bf v}$, 
such that near the singularity 
$$\chi = 2(N - 1) (1 - s)^{1/2} \ln r \equiv - 
{\beta \over 2\pi} \ln r, \eqno(21)$$     
and the flow corresponds to the line vortex, 
the strength of which according to (9) 
must be quantized $\beta = - 4\pi n = 4\pi m (1 - N)$. 
The velocity field near the center 
$${\bf v} = {-2n \over r}(- {y \over r}, {x \over r}),
\eqno(22)$$
is the gradient of multivalued function 
$\phi = -2n \arctan (y/x)$, 
which fixes the value of the phase of the 
wave function $\psi$. As is
well-known, functions $\chi$ and $\phi$ are 
conjugate harmonic functions, providing 
holomorphicity condition for the gauge potential 
near the center.
The above consideration shows that Chern-Simons 
soliton can be interpreted
as a planar vortex in the Madelung fluid, 
having form of a linear vortex 
near the rotation point.  This interpretation 
allows us to give the physical meaning for the 
vortex flow in the Madelung quantum liquid and 
answer the question posed in [19].    
\bigskip\bigskip
\centerline{3. EXPONENTIALLY LOCALIZED CHERN-SIMONS SOLITONS} 
\bigskip\bigskip\par
In previous section we considered self-duality reduction for
$s < 1$. However, no solution for $s > 1$ has been found. 
Now we show that in a special case
of hyperbolic energy surface 
in the dynamics of Bloch electrons 
under the influence of Chern-Simons and quantum potentials, 
the problem
admits an exact treatment also for $s > 1$. 
In the dynamics of Bloch electron in a 
solid a central role is played by 
the inverse effective mass tensor $1/m \rightarrow 
\partial^{2}/\partial {\bf k}^{2} E({\bf k})$, 
which is not necessary 
positive definite. When the Fermi surface is near 
the band boundary,
the sign of the mass in orthogonal direction is negative.  
In this 
hyperbolic case, due to strong Bragg reflection from 
the boundary of the 
band, the electron propagates along trajectories 
which are parallel to
the planes of the lattice, and $E({\bf k})$ has 
"saddle points", where
 the curvature of the surface may be positive 
in one direction and
negative in another. For two-dimensional motion, 
in simplest case
of the energy surface 
$E({\bf k}) = k^{2}_{1}/2m - k^{2}_{2}/2m $,
the effective mass matrix is $M^{-1} = diag (1/m, -1/m)$ 
and the Bloch
electron subject to the Lorentz force, 
imitated by Chern-Simons interaction, 
has a "dissipative" character [20], with the constant
magnetic field playing the role of the damping. 
Then, in Lagrangian (1) and equations of motion (2) 
we replace the 
positive space
metric $\eta_{ij} = diag (1,1)$ with the indefinite one
$\eta_{ij} = diag (1,-1)$ as follows 
$$
L = {\kappa \over 2}\epsilon^{\mu\nu\lambda}
A_{\mu}\partial_{\nu}A_{\lambda}
 + {i \over 2}(\bar\psi D_{0}\psi - \psi \bar D_{0}\bar \psi)
- \bar D_{a} \bar\psi  D^{a}\psi + 
s \partial_{a}|\psi| \partial^{a} |\psi|
+ g |\psi|^{4}, \eqno(23) $$
$$iD_{0}\psi + D_{a} D^{a}\psi + 2g |\psi|^{2}\psi = s 
{\partial_{a}\partial^{a} |\psi|
\over |\psi|}\psi, \eqno(24)$$
where $\partial_{a}\partial^{a} = 
\eta^{ab}\partial_{a}\partial_{b} =
\partial_{1}^{2} - \partial_{2}^{2}$. Now we 
rescale 
$$t = (s - 1)^{1/2} \tilde t,\,\,\,\,S = 
(s - 1)^{1/2}\tilde S, $$
$$A_{0} = (s - 1)\tilde A_{0}, \,\,\, {\bf A} =
(s - 1)^{1/2}\tilde {\bf A},\eqno(25)$$
and instead of $\psi = \exp (R - i S)$,  
introduce two real functions $Q^{\pm} = 
\exp (R \pm \tilde S)$,  satisfying the
system
$$-\tilde D_{\tilde 0}^{-} Q^{+} + 
\tilde D_{a}^{-}\tilde D^{a -}Q^{+}
- {2g \over s - 1} Q^{+}Q^{-}Q^{+} = 0, 
\eqno(26a)$$
$$\tilde D_{\tilde 0}^{+} Q^{-} + 
\tilde D_{a}^{+}\tilde D^{a +}Q^{-}
- {2g \over s - 1} Q^{+}Q^{-}Q^{-} = 0, \eqno(26b)$$
$$\partial_{1}\tilde A_{2} - \partial_{2}\tilde A_{1} =
{e \over \kappa (s - 1)^{1/2}}Q^{+}Q^{-},
\eqno(26c)$$
$$\partial_{\tilde 0}\tilde A_{j} - 
\partial_{j}\tilde A_{0} =
{e \over \kappa (s - 1)^{1/2}}
\epsilon_{jk} \eta_{kk}
(Q^{+} \tilde D_{k}^{+}Q^{-} - Q^{-} 
{\tilde D_{k}}^{-}Q^{+}),\,\, (j,k = 1,2),
\eqno(26d)$$ 
where $D^{\pm}_{\mu} = \partial_{\mu} \pm e A_{\mu}$.
This system represents the gauged version of the pair
time-reversal invariant diffusion equations 
with the reaction term proportional to $Q^{+}Q^{-}$. 
It is worth to note that switching on the 
Chern-Simons term leads to
switching off the time-reversal invariance 
without changing the geometry
of the classical trajectories. Moreover, 
instead of the local
phase transformations for the original model (23), 
in the system (26) the Chern-Simons 
gauge field corresponds to the local rescaling of $Q^{\pm}$.
For static configurations, when 
$\tilde A_{0} = - (e/\kappa(s - 1)^{1/2})
Q^{+}Q^{-}$, 
and 
$$ (s - 1)^{1/2} = - {g\kappa \over e^{2}},\eqno(27)$$
we have the self-duality equations
$$\tilde D^{-}_{-} Q^{+} = 0,
\,\,\,\,\,\tilde D^{+}_{+} Q^{-} = 0,
\eqno(28a)$$
$$[\tilde D^{+}_{+}, \tilde D^{+}_{-}] = -{2e^{2} 
\over \kappa (s - 1)^{1/2}}Q^{+}Q^{-} = -[\tilde D^{-}_{+}, 
\tilde D^{-}_{-}], \eqno(28b)$$
with $D^{a}_{\pm} = D^{a}_{1} \pm D^{a}_{2}$. Expressing 
$\tilde A_{\pm} = \mp (1/e)\partial_{\pm} \ln Q^{\mp}$ and 
substituting to (28b) for $\rho = Q^{+}Q^{-}$ we obtain the
Liouville equation
$$(\partial^{2}_{1} - \partial^{2}_{2}) 
\ln \rho = 2{e^{2} \over
\kappa (s - 1)^{1/2}} \rho. \eqno(29)$$
As is well-known this equation admits the 
general solution
$$\rho = {8 \over \alpha_{0}}{A'(x_{+})B'(x_{-}) 
\over [ A(x_{+}) + B(x_{-})]^{2}}, \eqno(30)$$
written in terms of two arbitrary functions 
$A(x_{+})$ and $ B(x_{-})$ of
$x_{+} = x_{1} + x_{2}$ and $x_{-} = x_{1} - x_{2} $ 
respectively,
where $\alpha_{0} = 2e^{2}/\kappa (s - 1)^{1/2}$. 
It has been considered
before as 1+1 dimensional evolution equation with 
$x$ coordinate considered
as a time variable. However, the known  
regular soliton solutions [21,22],
decay in all directions except the soliton world line, 
which leads to the
divergent integral for $\rho$, this is why they 
have no physical meaning in our 
problem. Instead of it   
we choose $A$ and $B$ functions in the form
$$A = (a + 1) \coth^{2k + 1}{\alpha \over 2} x_{+},\,\,
  B = (a - 1) \tanh^{2l + 1} {\beta \over 2} x_{-},
\eqno(31)$$     
then we get
$$\rho = {2^{1 - 2(k + l)} 
(1 - a^{2})(2k + 1)(2l + 1)\alpha\beta 
\sinh^{2k}\alpha x_{+} \sinh^{2l}\beta x_{-}
\over \alpha_{0} [(a + 1) \cosh^{2k + 1}{\alpha 
\over 2}x_{+} 
\cosh^{2l + 1}{\beta \over 2} x_{-} + 
(a - 1)\sinh^{2k + 1}{\alpha \over 2} x_{+} 
\sinh^{2l + 1}{ \beta \over 2} x_{-} ]^{2} }.$$
For parameter $a > 0$ this solution is nonsingular 
in the whole
$(x_{1},x_{2})$ plane.  Sign definiteness of 
$\rho$ requires $k, l$
to be integer, while regularity at 
the "light-cone", $x_{+} = 0$
or $x_{-} = 0$ is valid for $k \ge 0$, $l \ge 0$.
In contrast to the algebraic Chern-Simons solitons (8), 
solution (31) is exponentially decreasing in 
the plane for the "future"
and "past" null infinities as 
$e^{- \alpha |x|}$ or $e^{- \beta |x|}$,
while for the "time-like" and "space-like" 
infinities as  
$e^{- (\alpha + \beta)  |x|}$ 
(here without loose of generality we put  
$\alpha > 0$, $\beta > 0$). 
It is worth to note that our solutions
are related with exponentially localized planar solitons
($EL_{2}$) 
for Davey-Stewartson-I (DS-I) equation [23] 
(also known as dromions [24]). These
solutions have a rich integrable dynamics [25], 
and are reducible  
to the Liouville equation for $|\psi|$, and 
to the d'Alembert equation 
$\partial_{+}\partial_{-} \arg \psi = 0$
for $\arg \psi$, like in the DS-II case [26]. 
This reduction provides an integrable dynamics
for our exponentially localized Chern-Simons solitons [27].
Moreover, comparing to (8), having for $N >1$ the zero 
at the beginning of coordinates, the zeroes for solution (31) 
$\rho \sim x^{2k}_{+} x^{2l}_{-}$ 
are located along the "light cone"
$x_{+} = 0$, $x_{-} = 0$ 
for $k > 0$ and $l > 0$ correspondingly.
For $k = 0$ and  $l = 0$, 
we have exponentially localized soliton 
located at the beginning of coordinates, without zeroes 
and the rotation invariance at this point
(Fig.1). If only one of the $k, l$ values is vanishing,
then we have two soliton solution represented on Fig.2. 
and symmetrical 
under one of the light cone directions. When both
numbers do not vanish we get four-soliton solution, 
represented on Fig.3,
with zeroes along the whole light cone. 
For $k > 1$, $l > 1$, number
of solitons remains four, 
however the order of zeroes on the light cone is 
changing.
As is well known, zeroes of $\rho$, 
which produce singularities for
$\ln \rho$, lead to singularities of the gauge potential [3].
In the case (8) these singularities, 
corresponding to the Aharonov-Bohm
type potential, can be removed by 
fixing the phase of the wave function
at the singular point. So the phase becomes 
the angle variable on the plane with finite
range of values. For solution (31), 
the singularities of potential
${\bf A}$ are located along the light 
cone and can be compensated by
derivation of function $\tilde S$ as
$$\partial_{+}\tilde S = 2 k {1 \over x_{+}},\,\,\,\,
  \partial_{-}\tilde S = - 2 l {1 \over x_{-}},\eqno(32)$$
or 
$$\tilde S = {1 \over 2}\ln {x^{2k}_{+} 
\over x^{2l}_{-}}.\eqno(33)$$          
This allows us to define exact solutions 
of Eqs. (26) and original 
system (24)
$$\psi = \sqrt{\rho} \left({x^{l}_{-} \over x^{k}_{+}}
\right)^{i  \sqrt {s - 1}}. \eqno(34)$$     
For particular values $k = l$, the phase 
$\tilde S = k \ln |{x_{+} 
\over x_{-}}|$ defines the hyperbolic analog of the singular 
Aharonov-Bohm potential
$$a_{i} = {2 k \over e}\epsilon_{ij} {x_{j} 
\over x_{1}^{2} - x_{2}^{2}} =
 {k \over e} \epsilon_{ij} 
\eta_{jj}\partial_{j} \ln r^{2},\eqno(35)$$
where $r^{2} = x_{1}^{2} - x_{2}^{2}$ 
in the "time-like" quadrants
II $(x_{1} > 0)$ and IV $(x_{1} < 0)$ 
of the plane, while $r^{2} = x_{2}^{2} - x_{1}^{2}$ 
in the "space-like" quadrants
I $(x_{2} > 0)$ and III $(x_{2} < 0)$, parameterized as
$x_{2} = \pm r \cosh \theta$, $x_{1} = 
\pm r \sinh \theta$ in I and III,
and $x_{1} = \pm r \cosh \theta$, $x_{2} = 
\pm r \sinh \theta$ in II and IV,
correspondingly. Here 
$0 < r < \infty$, $-\infty < \theta < +\infty$. 
The potential (35) has singularities not only at 
the beginning of 
coordinates, as the Aharonov-Bohm potential, 
but also along the 
"light-cone", $x_{+} = 0$, $x_{-} = 0$ and 
may be presented as a
singular pure gauge
$$a_{i} = - {1 \over e} \partial_{i} \tilde S = 
-{2k \over e}
\partial_{i} \theta, \eqno(36)$$
where, $\tilde S = 2 k\theta$, and 
$$\theta = \cases{\tanh^{-1} {x_{1} \over x_{2}} 
& in I and III,\cr
           \tanh^{-1} {x_{2} \over x_{1}} & in 
II and IV. \cr}\eqno(37)$$
The magnetic flux associated with soliton solutions (23)
has the form
$$\int \tilde B d^{2}x = {e \over
\kappa (s - 1)^{1/2}}\int \rho d^{2}x = {2 \over e} 
\ln a,\eqno(38)$$
independent of $k$ and $l$, and is not quantized. 
Moreover, the phase of $\psi$ includes 
the hyperbolic rotation angle
$\theta$ which is valued on the 
whole real line. So no restriction of 
single-valuednees arises, and
continual 
parameters $s$ and $\kappa$ must be 
restricted only by the relation (27).
 \par 
At the end of this section we represent 
another problem related to our  self-dual system (28).
Namely, the hyperbolic self-duality equations (28) 
are equivalent to $SO(2,1)/O(1,1)$ self-dual $\sigma$
model:
$$\partial_{1}{\bf s} - {\bf s}\wedge\partial_{2}{\bf s} = 0, 
\eqno(39)$$
written in the tangent space representation for moving frame
$$D^{\mp}_{\mu} {\bf n}_{\pm} = 
\mp 2 \sqrt{{\alpha_{0} \over 2}}
Q^{\pm}_{\mu} {\bf s},\eqno(40a)$$
$$\partial_{\mu} {\bf s} = 
(-\sqrt{{\alpha_{0} \over 2}})
(Q^{+}_{\mu} {\bf n}_{-} - Q^{-}_{\mu}{\bf n}_{+}),\eqno(40b)
$$
to the one sheet hyperboloid ${\bf s}^{2} = 
- s^{2}_{1} + s^{2}_{2} - s^{2}_{3} = -1$ 
as the constraints $Q^{+}_{-} = 0$, $Q^{-}_{+} = 0$,
with the following  identification $Q^{+}_{+} 
\equiv Q^{+}$, $Q^{-}_{-} \equiv Q^{-}$. 
In terms of the stereographic projection of the  hyperboloid 
$$S_{\pm} = {2\xi_{\pm} \over 1 + \xi_{+}\xi_{-}},\,\,\,
  S_{3} = {1 - \xi_{+}\xi_{-} 
\over 1 + \xi_{+}\xi_{-}},\eqno(41)$$
Eqs. (39) are just the chirality conditions
$$\partial_{+}\xi_{+} = 0, \partial_{-}\xi_{-} = 0,\eqno(42)$$
having the general solution $\xi_{+} = 
\xi_{+} (x_{-})$, $\xi_{-} = 
\xi_{-} (x_{+})$. These functions correspond 
to the general solution of
the Liouville equation (30) with identification
$\xi_{-} = A^{-1}(x_{+})$, $\xi_{+} = B(x_{-})$. 
Then, our vortex
configurations (31) generate solution of (39),(42) with 
$$\xi_{-} = (a + 1)^{-1} \tanh^{2k + 1}
{\alpha \over 2} x_{+},\,\,\,\,\,
  \xi_{+} = (a - 1) \tanh^{2l + 1} {\beta \over 2} 
x_{-}.\eqno(43)$$ 
which for $a > 0$ are regular everywhere 
on the plane with $S_{3} > 0$. The last condition 
means that the solutions for $a > 0$ are non-topological. 
To have topologically nontrivial configurations 
we need to consider $a < 0$ case, which would 
produces singularity for $\rho$. 
For the regular solution ${\bf s}$ with $k > 0$ or(and) $l > 0$ 
the plateau along the light cone  
appears, leading to the zeroes for $\rho$ in (31).
\bigskip\bigskip
\centerline{4. CONCLUSIONS}
\bigskip\bigskip\par
In conclusions we stress that two long-range 
interactions considered in the present paper,  
the Chern-Simons gauge interaction 
introduced to
physics by S. Deser, R. Jackiw and S. Templeton [1]
and 
the quantum potential introduced by L. de Broglie [6] and D. Bohm [7], are compatible 
in supporting static soliton solutions in 2+1 
dimensions, with arbitrary $N$, only when 
the coupling constants for both interactions are quantized.
\par
A further remark concerns the special case $s = 1$, when
the Madelung hydrodynamical formulation of quantum mechanics
becomes the Euler equation and the continuity one [15].
For a special form of the nonlinearity an additional higher
symmetry of the equations, related to the membrane theory,
has been described recently [30]. It would be interesting 
to extend this work to include a Chern-Simons gauge field,
in attempt to produce completely integrable model with infinite number of conservation laws.
\par Finally, we note existence of another 
integrable reduction of the models
(1) and (23), when the fields are 
independent on one of the space directions. In this
case for $s < 1$ the model reduces to the BF gauged NLS,
equivalent to NLS [28], and for $s > 1$, 
to BF gauged reaction-diffusion analog
of NLS, equivalent to the reaction-diffusion system. 
The last one appears in
the Jackiw-Teitelboim gravity 
and admits dissipative analog of
solitons, called {\it dissipatons} 
and related to the black holes of the
model [29]. The interaction of dissipatons 
show the resonance character [16].
\bigskip\bigskip
\centerline{ACKNOWLEDGMENTS}
\bigskip\bigskip\par
The author are grateful to Roman Jackiw for his interest to
this paper and useful comments.
\par\bigskip\bigskip
\centerline{ REFERENCES}
\bigskip
\par
1. Deser, S., Jackiw, R. and Templeton, S., 
Topologically massive gauge theories, {\it Ann. Phys. (NY)}, 1982, {\bf 140}, 372.
\par
2. Wilczek, F., {\it Fractional Statistics 
and Anyon Superconductivity},
World Scientific, Singapore, 1990.
\par 
3. Jackiw, R. and Pi, S.-Y., {\it Phys. Rev. Lett.}, 
1990 ,{\bf 64}, 2969;
Classical and quantal nonrelativistic Chern-Simons theory,
{it Phys. Rev. D}, 1990, {\bf 42}, 3500-3513.
\par
4. Dunne, G., {\it Self-dual Chern-Simons Theories},
Springer Verlag, Berlin, 1995.
\par
5. Pashaev, O. K., Integrable 
Chern-Simons gauge field theory in 2 + 1 dimensions, 
{\it Mod. Phys. Lett. A}, 1996, {\bf 11}, 1713-1728.
\par
6. de Broglie, L., {\it  C.R. Acad. Sci.} 
(Paris), 1926, {\bf 183}, 447.
\par
7. Bohm, D., A Suggested Interpretation of the 
Quantum Theory in Terms of "Hidden" Variables.I, 
{\it Phys. Rev.}, 1952, {\bf 85}, 166-179.
\par
8. Weinberg, S., Testing Quantum Mechanics, 
{\it  Ann. Phys.}, 1989, {\bf 194}, 336-386.
\par
9. Doebner, H.-D. and Goldin, G. A., 
Properties of nonlinear Schr\"odinger equations 
associated with diffeomorphism group representations,
{\it J. Phys. A}, 1994,  {\bf 27}, 1771-1780, 
see references in  Doebner, H.-D., Goldin, G. A. 
and Nattermann, P., Gauge transformations in 
quantum mechanics and the unification of 
nonlinear Schr\"odinger equations, {\it J. Math. Phys.}, 1999,
{\bf 40}, 49-63. 
\par
10. Nelson, E., Derivation of 
Schr\"odinger equation from Newtonian mechanics, 
{\it Phys. Rev.}, 1966,  {\bf 150}, 1079-1085.
\par
11. Vigier, J.-P., Particular solutions of a 
non-linear Schr\"odinger equation carrying 
particle-like singularities represent possible 
models of de Broglie's double solution theory, 
{\it Phys. Lett.A}, 1989, {\bf 135}, 99-105. 
\par
12. Guerra, F. and Pusterla, M., A Nonlinear 
Schr\"odinger Equation and Its Relativistic 
Generalization from Basic Principles, 
{\it Lett. Nuovo Cimento}, 1982, {\bf 34}, 351-356;
\par
13. Smolin, L., Quantum fluctuations and inertia, 
{\it Phys. Lett.A}, 1986, {\bf A 113}, 408-412.
\par
14. Bertolami, O., Nonlinear corrections to 
quantum mechanics from quantum gravity, 
{\it Phys. Lett. A}, 1991, {\bf 154}, 225-229.
\par
15. Sabatier, P. C., Multidimensional nonlinear 
Schr\"odinger equations with exponentially 
confined solutions, {\it Inverse Problems}, 1990,  
{\bf 6}, L47-L53; Auberson, G.
and Sabatier, P. C., On a class of 
homogeneous nonlinear Schr\"odinger
equations, {\it  J. Math. Phys.}, 
1994, {\bf 35}, 4028-4040.
\par
16. Pashaev, O. K. and Lee, J.-H., 
Resonance NLS solitons as black holes in Madelung fluid, 
hep-th/9810139.
\par
17. Bateman, H., {\it Partial differential 
equations of mathematical physics}, Dover Pub., 
New York, 1944.
\par
18. Feynman, R. P., Leighton, R. B. and Sands, M. L., 
{\it Feynman lectures on physics}, v. 3, 
Addison-Weslay, Redwood City, CA,
1989.
\par 
19. Kozlov, V. V., {\it Obshaya teoriya vikhrey, 
(General vortex theory)},
Udmurtskiy Universitet Pub., Izhevsk, 1998 (in Russian). 
\par
20. Blazone, M. , Graziano, E., Pashaev, O. K. 
and Vitiello, G.
Dissipation and Topologically Massive Gauge 
Theories in the Pseudo-Euclidean Plane, 
{\it Annals of Physics} (NY), 1996, {\bf 252}, 115-132.
\par
21. Andreev, V. A., Application of the 
inverse scattering method to the equation 
$\sigma_{xt} = e^{\sigma}$, {\it Theor.  Math. Phys.}, 
1976, {\bf 29}, 213-220.
\par
22. Barbashov, B. M., Nesterenko, V. V. and Chervyakov, A. M.,
Solitons of some geometrical field theories, 
{\it Theor. Math. Phys.},
1979, {\bf 40},  15-27.
\par
23. Boiti, M., Leon, J., Martina, L. and Pempinelli, F., 
Scattering of localized solitons in the plane, 
{\it Phys. Lett. A}, 1988, {\bf 132}, 432-439.
\par
24. Fokas, A. S. and Santini, P. M., Dromions 
and a boundary value problem for 
Davey-Stewartson 1 equation, {\it Physica D}, 
1990, {bf 44}, 99-130.
\par
25. Boiti, M., Martina, L., Pashaev, O. K. 
and Pempinelli, F,
Dynamics of multidimensional solitons, 
{\it Phys. Lett. A}, 1991,{\bf 160}, 55-63.
\par
26. Arkadiev, V. A., Pogrebkov, A. K. and 
Polivanov, M. C., Closed string-like 
solutions of the Davey-Stewartson equation, 
{\it Inverse Problems}, 1989, {\bf 5}, L1-L6.
\par 
27. Lee, J.-H. and Pashaev, O. K. , in preparation.
\par 
28. Lee, J.-H, and Pashaev, O. K., Moving frames 
hierarchy and BF theory, {\it J. Math. Phys.}, 
1998, {\bf 39}, 102-123.
\par
29. Martina, L., Pashaev, O. K.  and Soliani, G., 
Integrable dissipative structures 
in the gauge theory of gravity, {\it Class. Quantum Grav.},
1997, {\bf 14}, 3179-3186; Bright solitons 
as black holes, {\it Phys. Rev. D}, 1998, {\bf 58}, 084025. 
\par
30. Bazeia, D., Jackiw, R., Nonlinear realization of a dynamical Poincare symmetry by a field-dependent diffeomorphism, {\it Ann. Phys. (NY)}, 1998, {\bf 270},
246-259; 
Jackiw, R. and Polychronakos, A. P., Fluid Dynamical Profiles and Constant of Motion from d-Branes, hep-th/9902024.
\filbreak
\bigskip\bigskip
\centerline{\bf Figures}
\bigskip\bigskip
{\bf Fig. 1.} Contour and 3D plot of one 
soliton solution (k = 0, l = 0) for a = 0.7 in (x,y) plane.
\bigskip\bigskip
{\bf Fig. 2.} Contour and 3D plot  
of two soliton solution (k = 1, l = 0)
for a = 0.7 in (x,y) plane.
\bigskip\bigskip
{\bf Fig. 3.} Contour and 3D plot 
of four soliton solution (k = 1, l = 1) 
for a = 0.7 in (x,y) plane.
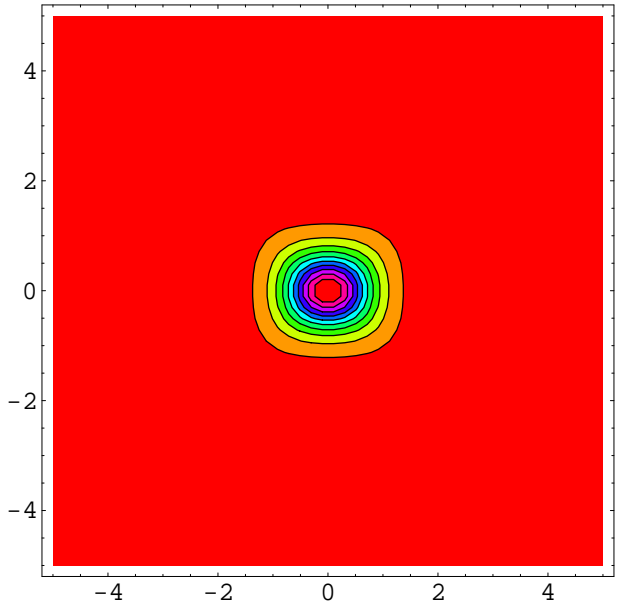
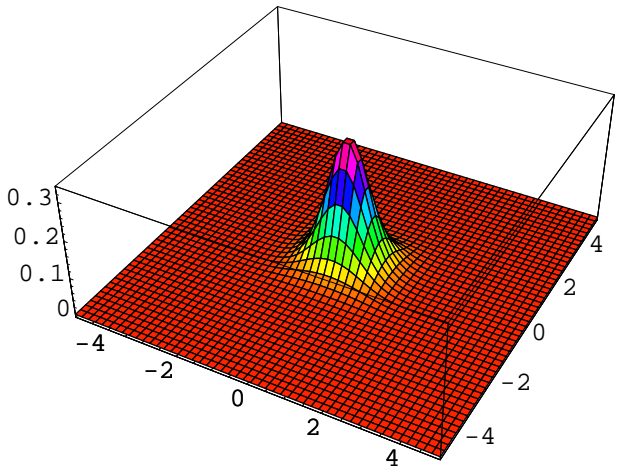
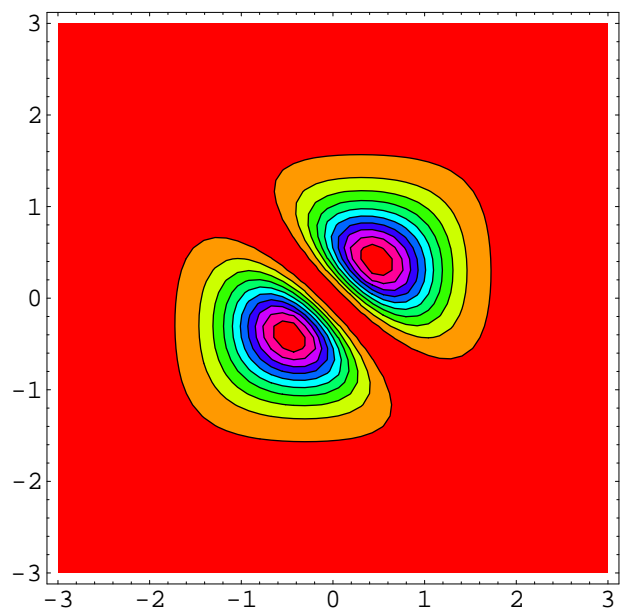
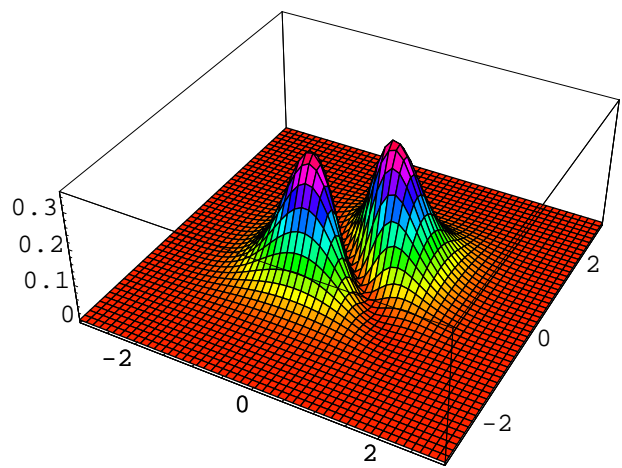
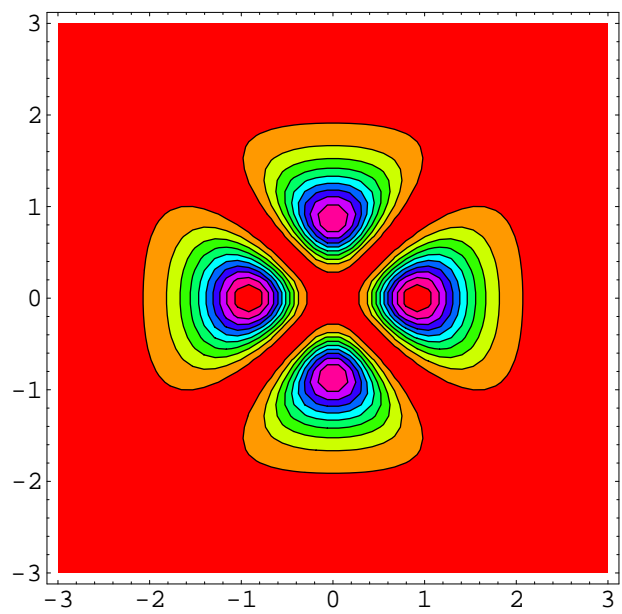
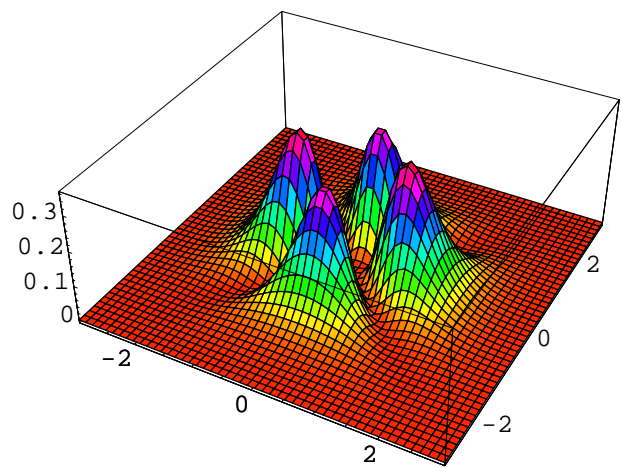
\end